\documentclass[manuscript]{aastex}

\slugcomment{Published in ApJL}

\shorttitle{Abundances of the Old Solar Twin HIP 102152}
\shortauthors{Monroe et al.}

\begin{document}

\title{High Precision Abundances of the Old Solar Twin HIP 102152:  Insights on Li Depletion from the Oldest Sun\altaffilmark{*}}

\author{TalaWanda R. Monroe\altaffilmark{1}}

\email{tmonroe@usp.br}

\author{Jorge Mel\'{e}ndez\altaffilmark{1}}
\author{Iv\'an Ram{\'{\i}}rez\altaffilmark{2}}
\author{David Yong\altaffilmark{3}}
\author{Maria Bergemann\altaffilmark{4}}
\author{Martin Asplund\altaffilmark{3}}
\author{Jacob Bean\altaffilmark{5}}
\author{Megan Bedell\altaffilmark{5}}
\author{Marcelo Tucci Maia\altaffilmark{1}}
\author{Karin Lind\altaffilmark{6}}
\author{Alan Alves-Brito\altaffilmark{3}}
\author{Luca Casagrande\altaffilmark{3}}
\author{Matthieu Castro\altaffilmark{7}}
\author{Jos\'e--Dias do Nascimento\altaffilmark{7}}
\author{Michael Bazot\altaffilmark{8}}
\author{Fabr\'{i}cio C. Freitas\altaffilmark{1}}

\altaffiltext{1}{Departamento de Astronomia do IAG/USP, Universidade de S{\~a}o Paulo, Rua do Mat{\~a}o 1226, Cidade Universit{\'a}ria, 05508-900 S{\~a}o Paulo, SP, Brasil}
\altaffiltext{2}{McDonald Observatory, The University of Texas at Austin,
Austin, TX 78712, USA}
\altaffiltext{3}{Research School of Astronomy and Astrophysics,
The Australian National University, Cotter Road, Weston, ACT 2611, Australia}
\altaffiltext{4}{Max Planck Institute for Astrophysics, Postfach 1317, 85741
Garching, Germany}
\altaffiltext{5}{Department of Astronomy and Astrophysics, University of Chicago,
5640 S. Ellis Ave, Chicago, IL 60637, USA}
\altaffiltext{6}{Institute of Astronomy, University of Cambridge, Madingley Road,
Cambridge, CB3 0HA, UK}
\altaffiltext{7}{Departamento de F{\'{\i}}sica Te\'orica e Experimental,
Universidade Federal do Rio Grande do Norte, 59072-970 Natal, RN, Brazil}
\altaffiltext{8}{Centro de Astrof\'{i}sica da Universidade do Porto, Rua das
Estrelas, 4150-762 Porto, Portugal}
\altaffiltext{*}{Based on observations obtained at the
European Southern Observatory (observing programs 083.D-0871 and 188.C-0265).}

\begin{abstract}

We present the first detailed chemical abundance analysis of the old 8.2 Gyr solar twin, HIP~102152.  We derive differential abundances of 21 elements relative to the Sun with precisions as high as 0.004 dex ($\lesssim$1\%), using ultra high-resolution (R = 110,000), high S/N UVES spectra obtained on the 8.2-m Very Large Telescope.  Our determined metallicity of HIP~102152 is [Fe/H] = -0.013 $\pm$ 0.004.  The atmospheric parameters of the star were determined to be 54 K cooler than the Sun, 0.09 dex lower in surface gravity, and a microturbulence identical to our derived solar value.  Elemental abundance ratios examined vs. dust condensation temperature reveal a solar abundance pattern for this star, in contrast to most solar twins.  The abundance pattern of HIP~102152 appears to be the most similar to solar of any known solar twin.  Abundances of the younger, 2.9 Gyr solar twin, 18~Sco, were also determined from UVES spectra to serve as a comparison for HIP~102152.  The solar chemical pattern of HIP~102152 makes it a potential candidate to host terrestrial planets, which is reinforced by the lack of giant planets in its terrestrial planet region.  The following non-local thermodynamic equilibrium Li abundances were obtained for HIP~102152, 18~Sco, and the Sun:  log $\epsilon$ (Li) = 0.48 $\pm$ 0.07, 1.62 $\pm$ 0.02, and 1.07 $\pm$ 0.02, respectively.  The Li abundance of HIP~102152 is the lowest reported to date for a solar twin, and allows us to consider an emerging, tightly constrained Li-age trend for solar twin stars.

\end{abstract}


\section{Introduction}

When compared to most stars with similar stellar parameters, the Sun displays an anomalous chemical abundance pattern.  To date, high-resolution spectroscopic chemical abundance analyses offer that only $\sim$15\% of solar type stars have abundance patterns similar to the Sun (Mel\'{e}ndez et al. 2009; Ram{\'{\i}}rez et al. 2009).  The disparity between the Sun and other solar twins has been demonstrated in a number of studies (e.g., Mel\'{e}ndez et al. 2009; Ram{\'{\i}}rez et al. 2010, 2011; Gonzalez et al. 2010a; Schuler et al. 2011), and is possibly due to the formation of terrestrial planets.  Planetesimal and terrestrial planet formation may have imprinted a signature in the solar composition by locking-up refractory elements in the surrounding disk, and leaving behind material low in refractory elements to later accrete onto the Sun.  

In this letter, we present elemental abundance ratios as a function of condensation temperature for the $\sim$8.2 Gyr old solar twin HIP~102152 (HD~197027).  This 1~M$_{\sun}$ star (0.97~M$_{\sun}$; current study) that is about to leave the main sequence is of interest not only because it has an abundance pattern similar to the Sun, thus making it an excellent older solar analog for comparative studies, but also because its solar abundance pattern may suggest that this star is a candidate host for terrestrial planets.  We also include abundances of the younger 2.9 Gyr solar twin 18~Sco and examine systematic differences between refractory and volatile elements in this representative solar twin and the older HIP~102152.  Additionally, we report Li abundances for the two stars and examine the tight Li-age depletion trend of solar twins studied at high-resolution, bringing further insights to the debate of whether low lithium is a signature of planets, or just an effect of secular stellar depletion (e.g., Takeda et al. 2007;  Takeda \& Tajitsu 2009; Ram{\'{\i}}rez et al. 2012; Baumann et al. 2010; Ghezzi et al. 2010; Gonzalez et al. 2010b; Israelian et al. 2009).

\section{Spectroscopic Observations and Reductions}

Spectra of HIP~102152 and 18~Sco were obtained with the Ultra-violet and Visible Echelle Spectrograph (UVES; Dekker et al. 2000) on the 8.2-m Very Large Telescope (VLT) at the ESO Paranal Observatory.  Observations were carried out on 30 August 2009, with air masses of 1.1--1.5 for HIP~102152 and 1.6--1.7 for 18~Sco.  Multiple exposures were obtained at three spectrograph configurations (six exposures per configuration) for each star to achieve redundancy in the observations around the Li doublet at $\lambda$6708\AA, with individual exposure times of 10--20 min for HIP102152 and 45--90 s for the brighter 18~Sco.  Spectral coverage included 3060--3870\AA, 4800--5770\AA, 5850--8200\AA, and 8430--10200\AA, with incomplete coverage in the near-infrared due to gaps between echelle orders.  The spectral resolving powers were R($\lambda$/$\delta\lambda$) = 65,000 at $\lambda$3060--3870\AA, and R = 110,000 for $\lambda$4800--10200\AA.  The asteroid Juno was also observed with the same spectrograph setups, and served as the solar reference in our differential analysis.

Standard reductions were carried out in IRAF, including bias subtraction, flat fielding, wavelength calibrations, spectral order extractions, and spectra co-additions.  Continuum normalization was performed on the combined one-dimensional spectra in IDL.  Final composite spectra had signal-to-noise (S/N) ratios that varied across the spectral orders, but had a typical S/N$\sim$500 pixel$^{-1}$. It should be noted that our spectrograph setups allowed us to achieve S/N$\sim$1000 pixel$^{-1}$ in the continuum regions about the Li feature at $\lambda$6708\AA.

\section{Analysis}
\subsection{Chemical Abundances}

Chemical abundances of most species were determined similarly to the methods outlined in Mel\'{e}ndez et al. (2012).  Equivalent width analyses were performed using the \emph{abfind} driver in the local thermodynamic equilibrium (LTE) line analysis code MOOG (2002 version; Sneden 1973) and Kurucz (ATLAS9) model atmospheres, without convective overshooting (Castelli \& Kurucz 2004).    

LTE abundances were determined for the following Z$\leq$30 elements in our analysis: C, N, O, Na, Mg, Al, Si, S, K, Ca, Sc, Ti, V, Cr, Mn, Fe, Co, Ni, Cu, and Zn; and are included in Table \ref{tab1}.  The line list was mostly taken from Mel\'{e}ndez et al. (2012).  Equivalent width (EW) measurements were first made with the automated code ARES (Sousa et al. 2007).  Lines were measured by hand in both the star and Sun with the IRAF task \emph{splot} when the EW $<$ 10 m\AA, or if the species had five or fewer lines.  Additionally, lines that were found to be outliers from the mean abundance of a species were also measured by hand.  Measurements were performed by fitting a single Gaussian profile to the line profile, or deblending multiple Gaussian profiles when necessary.  To aid in continuum placement, the observed solar spectrum was overplotted on each stellar line.  The measurements for the star and Sun were then made in an identical manner, using the same regions of continuum.

All elemental abundances were computed line-by-line differentially with respect to the Sun, using the same line list.  A strictly differential approach was taken to decrease the additional line-to-line scatter that can arise from uncertainties in the adopted transition probabilities.  Also, the differential approach minimizes systematic errors in the determinations of stellar parameters, as well as systematic uncertainties in the abundances due to deficiencies in the model atmospheres.  For the odd-Z elements of V, Mn, Co, and Cu, we took into account effects of hyperfine splitting (HFS).  Abundances of these elements were performed using our measured EWs and the \emph{blends} driver in MOOG, which allowed us to consider the numerous individual components and wavelength splittings of the measured lines.  The HFS line lists used were adopted from Mel\'{e}ndez et al. (2012).  Additionally, abundances of the Li resonance doublet feature at $\lambda$=6708\AA\ were determined using spectrum synthesis, and will be discussed below.  Differential NLTE corrections were shown to be very small in solar twins relative to the Sun ($\sim$0.001 dex) by Mel\'{e}ndez et al.; therefore, we did not take them into account in this study, except for Cr, Mn, and Co (e.g., Bergemann \& Cescutti 2010). The sensitivity of our differential chemical abundances to the choice of model atmosphere grid has been shown to be negligible (Mel\'{e}ndez et al.).

\subsection{Atmospheric Parameters}

Abundances were first determined for the Sun using the standard solar atmospheric parameters of T$_{\rm eff}$=5777 K and log g=4.44.  We determined our own value of the solar microturbulent velocity (v$_{t}$) by using the common spectroscopic technique of removing any trend between the abundances of Fe I lines and reduced equivalent width.  The resulting microturbulence was v$_{t}$ = 0.86 km s$^{-1}$, although this value was individually tweaked slightly for the analyses of both HIP~102152 and 18~Sco.

An initial EW analysis was performed for the stars using the above atmospheric parameters adopted for the Sun.  Effective temperatures (T$_{\rm eff}$) for the program stars were initially set at the solar value and then refined by employing excitation equilibrium on the differential abundances of the Fe I lines.  That is, all trends between the differential abundances of each line with excitation potential (EP) were removed.  Stellar microturbulences were refined by removing trends between the differential abundances and reduced EW.  Surface gravities (log g) were incrementally adjusted from solar to arrive at ionization balance between the average differential iron abundances derived from Fe I and Fe II lines.

Final atmospheric parameters for the stars are presented in Table \ref{tab1}.  We determined the following values for HIP~102152:  T$_{\rm eff}$ = 5723$\pm$5 K, log g = 4.35$\pm$0.02 dex, and v$_{t}$ = 0.86$\pm$0.01 km s$^{-1}$ ($\Delta$v$_{t}$ = 0.0 km s$^{-1}$ for v$_{t,\odot}$ = 0.86 km s$^{-1}$).  The differential stellar parameters between HIP~102152 and the Sun are therefore -54 K in T$_{\rm eff}$, -0.09 dex in log g, and 0.0 km s$^{-1}$ in microturbulence, and thus confirm that this older dwarf is excellent for comparisons to the Sun.  Our T$_{\rm eff}$ is in good agreement with previous spectroscopic determinations by the lower resolution studies by Ram{\'{\i}}rez et al. (2009), who reported a temperature of 5746$\pm$50 K, and Baumann et al. (2010), who report a T$_{\rm eff}$ of 5737$\pm$47 K.  The previous estimates of surface gravities are also in excellent agreement with our determinations:  log g = 4.40$\pm$0.07 (Ram{\'{\i}}rez et al.) and 4.35$\pm$0.06 (Baumann et al.).  Our atmospheric parameters for 18~Sco (T$_{\rm eff}$ = 5824$\pm$5 K, log g = 4.45$\pm$0.02 dex, and  v$_{t}$ = 0.89$\pm$0.01  km s$^{-1}$)\footnote{$\Delta$v$_{t}$ = 0.02 km s$^{-1}$ for v$_{t,\odot}$ = 0.87 km s$^{-1}$.} are also in agreement with independent determinations in Mel\'{e}ndez et al. (2012; 2013).  We estimate that the uncertainties in T$_{\rm eff}$, log g, microturbulence, and metallicity of the adopted model atmospheres are 5K, 0.02 dex, 0.01 km s$^{-1}$, and 0.01 dex, respectively.  These errors include consideration of measurement uncertainties and the degeneracies in the stellar parameters.  Abundance uncertainties for each element due to the errors in the stellar parameters are included in Table \ref{tab1}.  The last column in the table includes a total uncertainty for each element, obtained by adding in quadrature the uncertainties from the stellar parameters and observational standard error of the mean (SEM).  The cumulative error from the stellar parameters is 0.004 dex for [Fe/H].

\section{Solar Abundance Trend of HIP~102152}

The metallicities of HIP~102152 and 18~Sco were determined to be [Fe/H] = -0.013 $\pm$ 0.004 and 0.055 $\pm$ 0.004, respectively, as indicated in Table \ref{tab1}.  The solar metallicity of HIP~102152 and slightly supersolar abundance of 18~Sco were also reported in Baumann et al. (2010), who similarly found [Fe/H] = -0.010 $\pm$ 0.022 for HIP~102152 and [Fe/H] = 0.051 $\pm$ 0.020 for 18~Sco.

\begin{figure}
\plotone{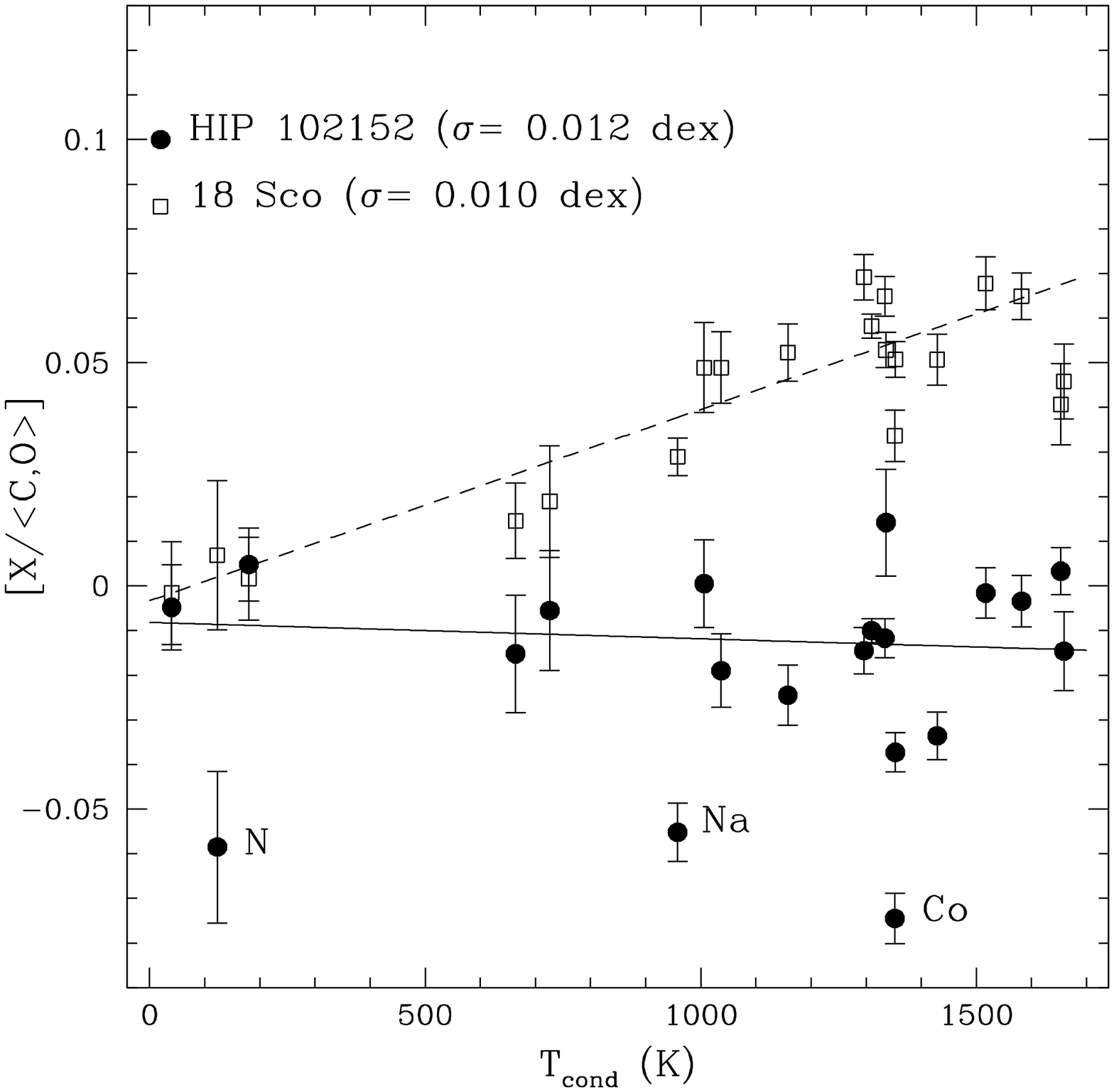}
\caption{Differential abundance ratios of HIP~102152 (filled circles) and 18~Sco (open squares) vs. dust condensation temperature.  Plotted abundances are relative to $<$C,O$>$.  Error bars include contributions from the observational uncertainties (SEM) and errors associated with the stellar atmospheric parameters.  Weighted linear fits are included for both stars (solid line:  HIP~102152; dashed line:  18~Sco).  Three outliers are labelled.}
\label{f1}
\end{figure}

Elemental abundance ratios ([X/$<$C,O$>$]) of species with Z$\leq$30 are plotted vs. equilibrium condensation temperature (T$_{C}$) for HIP~102152 and 18~Sco in Figure \ref{f1}.  The abundance ratios are relative to the average of the carbon and oxygen abundances ($<$C,O$>$), since those elements may be undepleted in the solar atmosphere (Mel\'{e}ndez et al. 2009, 2012), and allow us to visually compare volatile (T$_{C}$ $<$ 900 K) and refractory (T$_{C}$ $>$ 900 K) species easily.  The differential abundances of HIP~102152 display a pattern that is similar to solar, with a slope (solid line) consistent with zero (-3.7 $\times$ 10$^{-6}$ $\pm$ 9.9 $\times$ 10$^{-6}$ K$^{-1}$) and an element-to-element scatter of 0.012 dex, after excluding three outliers (N, Na, and Co) that have abundances significantly lower than the rest.  Given the large age of HIP~102152, the low abundances of these three elements may be due to a lower contribution from Type Ia supernovae (Co; Figure~7 in Mel\'{e}ndez et al. 2012) and asymptotic giant branch (AGB) stars (N and Na).  We will re-examine these possible scenarios for our larger sample of solar twins of disparate ages.  Abundance ratios of neutron capture elements with Z$>$30 also appear roughly solar for the r-process elements, but the s-process elements appear deficient, which is consistent with the low N and Na abundances resulting from the lower AGB contribution.  These results will be presented elsewhere.

In contrast, our abundances of 18~Sco in Figure \ref{f1} agree with previous studies (e.g., Mel\'{e}ndez et al. 2012), which show the star has an abundance pattern similar to other solar twins.  That is, the refractory elements are enhanced relative to the volatile species, as illustrated by the dashed line.  A similar correlation has been demonstrated on larger samples of solar twins in a number of high-resolution studies (e.g., Mel\'{e}ndez et al. 2009; Ram{\'{\i}}rez et al. 2009; Schuler et al. 2011).  The element-to-element scatter for 18~Sco is low, at 0.010 dex.  As shown in Figure \ref{f1}, the abundances of refractory elements between HIP~102152 and 18~Sco differ by as much as 0.07 dex.

\section{Lithium Abundance vs. Stellar Age}

\begin{figure}
\plotone{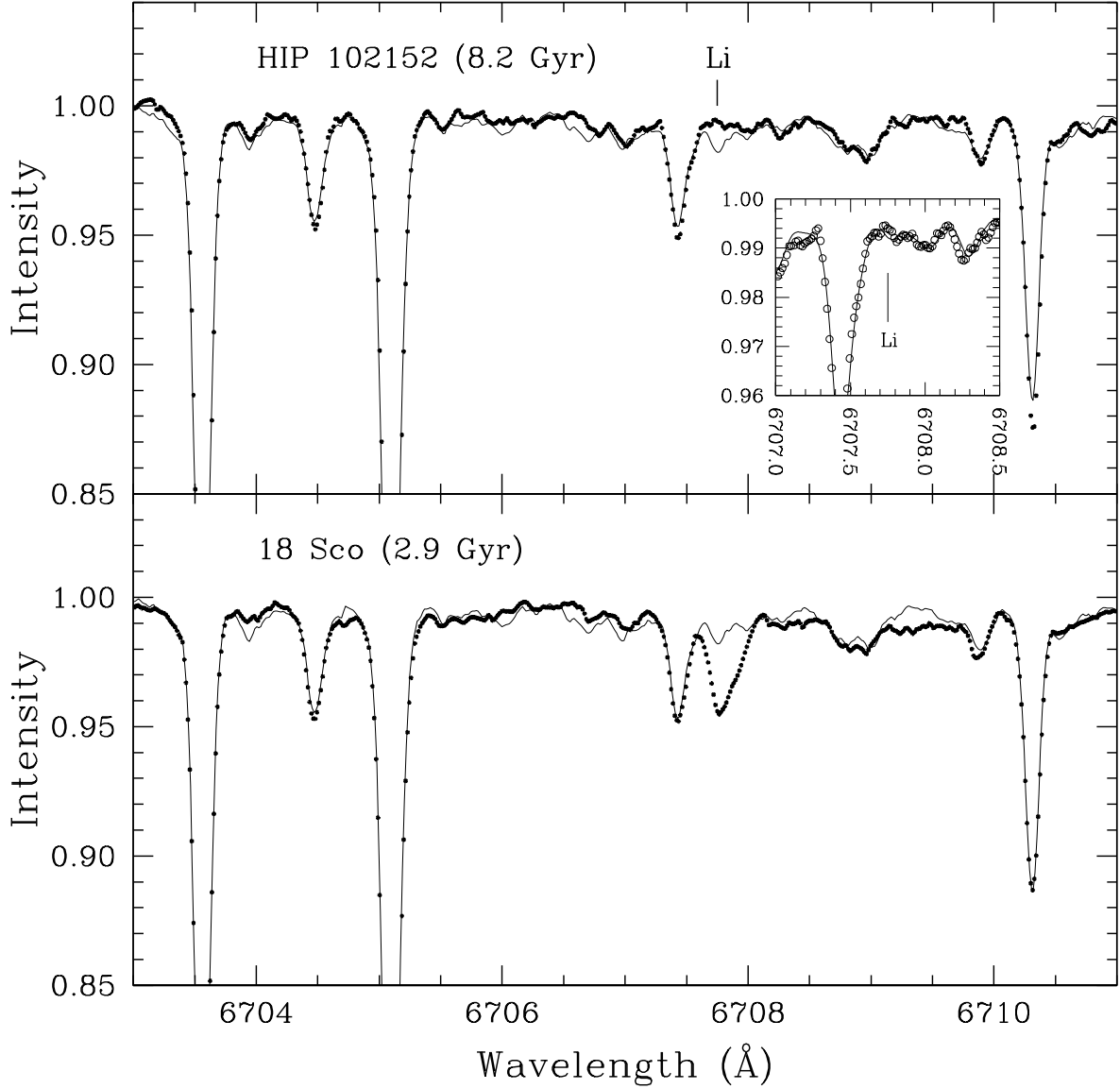}
\caption{The spectral region about the Li doublet at $\lambda$6708\AA\ of HIP~102152 and 18~Sco.  The stellar spectra are plotted with circles.  The solar spectrum is overlaid with a solid line in each panel, to visually demonstrate Li depletion with age.  The inset panel shows the synthetic spectrum (solid line) used to determine the Li abundance of HIP~102152.}
\label{f2}
\end{figure}

LTE lithium abundances were determined for HIP~102152, 18~Sco, and the Sun with spectrum synthesis using the \emph{synth} driver in MOOG.  We estimate the solar Li abundance to be log $\epsilon$ (Li) = 1.03 $\pm$ 0.02, in strong agreement with the reported value of log $\epsilon$ (Li) = 1.05 $\pm$ 0.10 by Asplund et al. (2009).  As shown in Figure \ref{f2}, the Li doublet is weak in the spectrum of HIP~102152, compared to the also depleted solar line.  The high S/N $\sim$1000 pixel$^{-1}$ in the region about the line allows us to place bounds on the  Li abundance of HIP~102152 from log $\epsilon$ (Li) $\sim$0.37--0.51 dex, with a best estimate of log $\epsilon$ (Li) = 0.43 $\pm$ 0.07.  In contrast, the line of the younger solar twin, 18~Sco, is much stronger.  We estimate the abundance of this star to be log $\epsilon$ (Li) = 1.58 $\pm$ 0.02.  NLTE Li abundance corrections were determined as in Lind et al. (2009), and estimated to be +0.04 dex for 18 Sco and the Sun, and +0.05 dex for HIP~102152.

\begin{figure}
\plotone{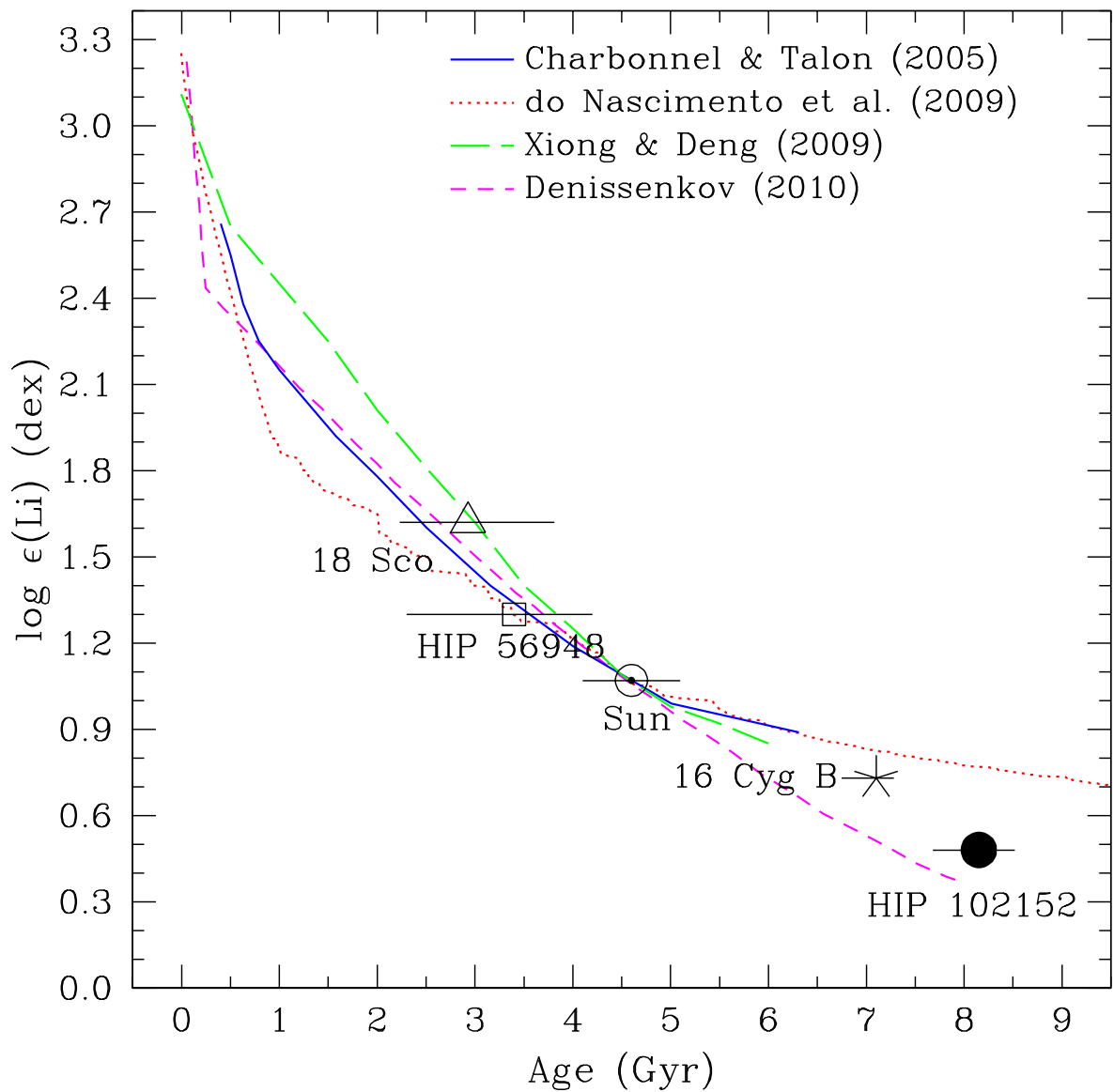}
\caption{NLTE Li abundance vs. isochronal age for the Sun and solar twins observed at high-resolution with high S/N.  Li abundance determinations for HIP~102152, 18~Sco, and the Sun were determined for this study.  Abundances of the other stars were taken from the literature, as noted in the text.  Error bars for Li are smaller than the data points.   Model predictions are also included and these models have been offset to intersect our derived solar Li abundance.}
\label{f3}
\end{figure}

Ischronal ages and masses were estimated for HIP~102152 and 18~Sco using Yonsei-Yale isochrones (Kim et al. 2002) based on their stellar parameters and associated errors using probability distribution functions as described in Mel\'{e}ndez et al. (2012).  In Figure \ref{f3} we present NLTE Li abundances vs. isochronal stellar age for a small group of solar twins studied at high-resolution (R$>$60,000) and high S/N ($>$400), along with four theoretical Li abundance models for a solar-mass, solar metallicity star (Charbonnel \& Talon 2005; do Nascimento et al. 2009; Xiong \& Deng 2009; and Denissenkov 2010).  From the published studies, we chose a model that best represented the Li abundance data.  In the interest of better characterizing and constraining the timescale of Li depletion models, we are concentrating on studies with high quality data for two reasons.  (1) For stars younger than the Sun, these constraints limit uncertainties in age, which have unfortunately impeded previous studies (e.g., Baumann et al. 2010), and (2) for stars older than the Sun, the high S/N ratios are crucial for obtaining accurate measurements using very weak Li lines (see Figure 2), which have restricted previous attempts (e.g., Takeda et al. 2007).  In Figure \ref{f3} Li abundances of HIP~102152, 18~Sco, and the Sun are from the current study, while the abundances and ages of HIP 56948 and 16~Cyg~B were taken from Mel\'{e}ndez et al. (2012) and Ram{\'{\i}}rez et al. (2011), respectively.  Age determinations for all the stars were determined using the same method.  Recently, do Nascimento et al. (2013) reported an old 6.7 Gyr solar twin with a low Li abundance of log $\epsilon$ (Li) $\leq$ 0.85 dex, but this value is unfortunately limited by low S/N observations.    As predicted and demonstrated previously, older stars show significantly more depletion than younger stars, with the abundance of HIP~102152 being the lowest Li abundance reported for a solar twin to date.  This low abundance along with the abundance of the widely separated binary star 16~Cyg~B may be useful in distinguishing between the depletion models, as well as helping to constrain the models for the upper end of the age range of solar-like main sequence stars, where the models start to diverge or are incomplete.  Overall the models represent the data well for the parameter space covered.  Given this surprisingly tight preliminary correlation, we are currently analyzing a larger set ($\sim$15) of solar twins of disparate ages observed with UVES to further fill in this diagram and hopefully confirm and strengthen the trend.

\section{Discussion \& Conclusions}

The solar twin HIP~102152 is notably similar to the Sun in mass, effective temperature, surface gravity, microturbulence, and metallicity.  The solar abundance pattern of HIP~102152 that we report is the closest to the solar pattern published to date, using high precision abundances.  The abundance ratios of HIP~102152 show no enhancements of refractory elements compared to volatile species, unlike other solar twins such as 18~Sco.  The abundances of the old star HIP~102152 are even closer to solar than the abundances of HIP 56948, a solar twin previously found to be the most similar to the Sun (Mel\'{e}ndez et al. 2012).  

\begin{figure}
\plotone{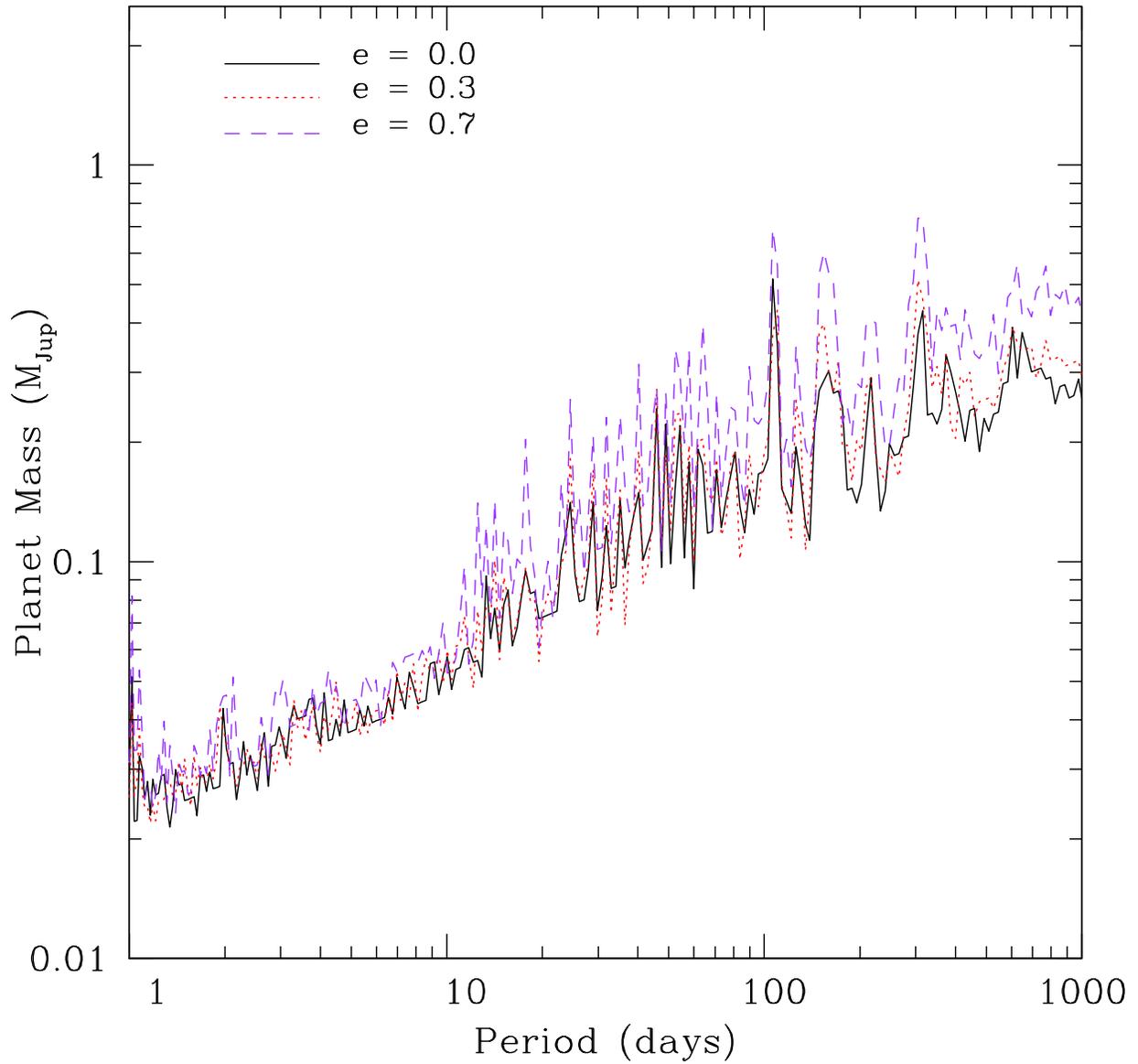}
\caption{Detection limits based on our HARPS data are shown for different eccentricities (e).  Planets above the curves are ruled out.}
\label{f4}
\end{figure}

The Sun appears deficient in refractory elements relative to volatile species, when compared to other solar twins (Mel\'{e}ndez et al. 2009).  Assuming the solar twins and the Sun formed under similar conditions, from gas of similar composition, then the solar refractory elements must be tied-up somewhere.  A reasonable inference by many authors is that the observed pattern is due to the accretion of dust-depleted gas onto the pre-main sequence Sun (e.g., Nordlund 2009; Chambers 2010).  Perhaps one of the more tantalizing mechanisms that has been suggested to produce dust-depleted gas in a natal disk is the formation of planetesimals and terrestrial planets, since the abundance patterns of the inner solar-system planets and meteorites are conversely enhanced in refractory elements (Ciesla 2008; Alexander et al. 2001).  If the solar abundance pattern is in fact a consequence of terrestrial planet formation, then other stars with abundance patterns similar to the Sun may be identified as potential rocky planet hosts.  In fact, the solar abundance pattern of HIP~102152 makes it a prime potential candidate host for terrestrial planets\footnote{Alternatively, stars born in a dense cluster may be subject to radiation pressure that may deplete dust in the circumstellar environment (e.g., {\"O}nehag et al. 2011).}.  This star is included in our long-term radial velocity monitoring survey of solar twins being carried out with HARPS on the ESO 3.6-m telescope; however, current observations show no evidence of velocity variations that may be attributed to giant planets within 2 AU (see Figure \ref{f4}).

While most stellar parameters of HIP~102152 are virtually identical to the Sun, the age of the star is significantly larger, by $\sim$3.6 Gyr.  The star's old age and physical similarities to the Sun make HIP~102152 an excellent older analog for the Sun.  In particular, this star is ideal for exploring the potential evolution of the Sun's lithium abundance, by probing the older end of the lithium-age trend both observationally and theoretically.  Lithium is severely depleted in HIP~102152, as predicted.  This star along with other solar twins firmly demonstrate that the Sun's Li has depleted as predicted for a star with its age, mass, and metallicity.  Previously, the apparently low solar Li abundance has been of concern for investigations that have compared the Sun to other solar analogs, with a broader range of stellar parameters (e.g., Lambert \& Reddy 2004).  However, when comparisons of the Sun's Li abundance are made strictly to other solar twins, the Sun appears normal (Mel\'{e}ndez et al. 2010).  The small amount of scatter found about the Li-age relation for the sample of stars considered here is encouraging, and worthy of followup.  This may be accomplished with the ongoing spectroscopic analysis of our larger sample of solar twins.  The high-resolution and high S/N of our UVES spectra are allowing us to determine more precise stellar parameters for the sample, and thus more precise isochronal ages.  It is expected that our remaining data set will address the longstanding problem of Li depletion and provide tight constraints for non-standard models of temporal Li evolution.

Some authors have claimed enhanced Li depletion in planet hosts, but Li-age trends have been largely ignored or not detected due to age determination uncertainties. Here we have derived reliable ages and highly-precise Li abundances, leading to a well-defined Li-age correlation that demonstrates the importance of the age bias in Li-planet studies.

\acknowledgments

TRM and JM acknowledge support from FAPESP (2010/19810-4 and 2012/24392-2).  IR's work was performed under contract with Caltech, funded by the NASA Sagan Fellowship Program.  We thank Bengt Gustafsson and Stefan Dreizler for useful comments.

{\it Facility:}\facility{VLT:Kueyen (UVES)}.

\clearpage

\clearpage

\clearpage

\clearpage

\clearpage

\tablecolumns{10}
\tablewidth{0pt}

\begin{deluxetable}{llllllllll}
\tabletypesize{\scriptsize}
\tablecaption{Stellar Abundances [X/H] and Errors \label{tab1}}

\tablehead{
\colhead{Element}      &
\colhead{[X/H]}      &
\colhead{SEM}      &
\colhead{N}          &        
\colhead{$\Delta$T$_{\rm eff}$}          &
\colhead{$\Delta$log g}          &
\colhead{$\Delta$v$_{t}$}          &
\colhead{$\Delta$[Fe/H]}          &
\colhead{$\Delta$Parameters}          &
\colhead{Total Unc.}                 \\

\colhead{}      &
\colhead{}      &
\colhead{}      &
\colhead{}          &        
\colhead{(+5 K)}          &
\colhead{(+0.02 dex)}          &
\colhead{(+0.01 km s$^{-1}$)}          &
\colhead{(+0.01 dex)}          &
\colhead{}          &
\colhead{}

}

\startdata

\hline
\multicolumn{10}{l}{HIP 102152  (5723 $\pm$ 5 K,  log g = 4.35 $\pm$ 0.02,  $\Delta$v$_{t}$ = 0.00 $\pm$ 0.01 km s$^{-1}$,  log $\epsilon$ (Li) = 0.48 $\pm$ 0.07)} \\
\hline
C	&	-0.006	&	0.006	&	4	&	-0.005	&	0.005	&	0.000	&	-0.001	&	0.007	&	0.010	\\
N	&	-0.060	&	0.013	&	1	&	-0.005	&	-0.001	&	-0.003	&	-0.009	&	0.011	&	0.017	\\
O	&	0.003	&	0.004	&	3	&	-0.007	&	0.002	&	-0.001	&	0.000	&	0.007	&	0.008	\\
Na	&	-0.057	&	0.006	&	3	&	0.002	&	0.000	&	0.000	&	0.001	&	0.003	&	0.007	\\
Mg	&	0.013	&	0.012	&	3	&	0.002	&	-0.002	&	-0.001	&	0.001	&	0.003	&	0.012	\\
Al	&	0.002	&	0.005	&	6	&	0.002	&	-0.002	&	0.000	&	0.001	&	0.003	&	0.005	\\
Si	&	-0.012	&	0.002	&	13	&	0.000	&	0.001	&	-0.001	&	0.001	&	0.002	&	0.003	\\
S	&	-0.017	&	0.012	&	6	&	-0.004	&	0.004	&	0.000	&	-0.001	&	0.006	&	0.013	\\
K	&	-0.001	&	0.003	&	1	&	0.004	&	-0.008	&	-0.002	&	0.003	&	0.009	&	0.010	\\
Ca	&	-0.003	&	0.002	&	10	&	0.003	&	-0.003	&	-0.002	&	0.002	&	0.005	&	0.006	\\
Sc	&	-0.016	&	0.004	&	9	&	0.000	&	0.007	&	-0.002	&	0.003	&	0.008	&	0.009	\\
Ti	&	-0.005	&	0.003	&	21	&	0.005	&	0.000	&	-0.001	&	0.001	&	0.005	&	0.006	\\
V	&	-0.035	&	0.001	&	9	&	0.005	&	0.001	&	-0.001	&	0.001	&	0.005	&	0.005	\\
Cr	&	 -0.016\tablenotemark{a}	&	0.002	&	16	&	0.004	&	-0.002	&	-0.002	&	0.002	&	0.005	&	0.005	\\
Mn	&	 -0.026\tablenotemark{a}	&	0.004	&	5	&	0.003	&	-0.003	&	-0.002	&	0.002	&	0.005	&	0.007	\\
Fe	&	-0.013	&	0.001	&	87	&	0.003	&	-0.001	&	-0.002	&	0.002	&	0.004	&	0.004	\\
Co	&	 -0.076\tablenotemark{a}	&	0.003	&	8	&	0.004	&	0.002	&	-0.001	&	0.002	&	0.005	&	0.006	\\
Ni	&	-0.039	&	0.002	&	17	&	0.003	&	0.000	&	-0.002	&	0.002	&	0.004	&	0.004	\\
Cu	&	-0.021	&	0.007	&	4	&	0.003	&	0.000	&	-0.002	&	0.002	&	0.004	&	0.008	\\
Zn	&	-0.007	&	0.013	&	2	&	-0.001	&	0.002	&	-0.002	&	0.002	&	0.003	&	0.013	\\

\hline
\multicolumn{10}{l}{18 Sco (5824 $\pm$ 5 K, log g = 4.45 $\pm$ 0.02,  $\Delta$v$_{t}$ = 0.02 $\pm$ 0.01 km s$^{-1}$, log $\epsilon$ (Li) = 1.62 $\pm$ 0.02) } \\
\hline
C	&	-0.012	&	0.009	&	4	&	\nodata	&	\nodata	&	\nodata	&	\nodata	&	\nodata	&	0.012	\\
N	&	-0.003	&	0.013	&	1	&	\nodata	&	\nodata	&	\nodata	&	\nodata	&	\nodata	&	0.017	\\
O	&	-0.008	&	0.006	&	3	&	\nodata	&	\nodata	&	\nodata	&	\nodata	&	\nodata	&	0.009	\\
Na	&	0.019	&	0.003	&	3	&	\nodata	&	\nodata	&	\nodata	&	\nodata	&	\nodata	&	0.004	\\
Mg	&	0.043	&	0.002	&	1	&	\nodata	&	\nodata	&	\nodata	&	\nodata	&	\nodata	&	0.004	\\
Al	&	0.031	&	0.009	&	6	&	\nodata	&	\nodata	&	\nodata	&	\nodata	&	\nodata	&	0.009	\\
Si	&	0.048	&	0.002	&	13	&	\nodata	&	\nodata	&	\nodata	&	\nodata	&	\nodata	&	0.003	\\
S	&	0.005	&	0.006	&	6	&	\nodata	&	\nodata	&	\nodata	&	\nodata	&	\nodata	&	0.008	\\
K	&	0.039	&	0.004	&	1	&	\nodata	&	\nodata	&	\nodata	&	\nodata	&	\nodata	&	0.010	\\
Ca	&	0.058	&	0.003	&	9	&	\nodata	&	\nodata	&	\nodata	&	\nodata	&	\nodata	&	0.006	\\
Sc	&	0.036	&	0.002	&	9	&	\nodata	&	\nodata	&	\nodata	&	\nodata	&	\nodata	&	0.008	\\
Ti	&	0.055	&	0.002	&	21	&	\nodata	&	\nodata	&	\nodata	&	\nodata	&	\nodata	&	0.005	\\
V	&	0.041	&	0.002	&	9	&	\nodata	&	\nodata	&	\nodata	&	\nodata	&	\nodata	&	0.006	\\
Cr	&	0.059\tablenotemark{a}	&	0.001	&	16	&	\nodata	&	\nodata	&	\nodata	&	\nodata	&	\nodata	&	0.005	\\
Mn	&	0.042\tablenotemark{a}	&	0.004	&	5	&	\nodata	&	\nodata	&	\nodata	&	\nodata	&	\nodata	&	0.006	\\
Fe	&	0.055	&	0.001	&	85	&	\nodata	&	\nodata	&	\nodata	&	\nodata	&	\nodata	&	0.004	\\
Co	&	0.024\tablenotemark{a}	&	0.003	&	8	&	\nodata	&	\nodata	&	\nodata	&	\nodata	&	\nodata	&	0.006	\\
Ni	&	0.041	&	0.001	&	17	&	\nodata	&	\nodata	&	\nodata	&	\nodata	&	\nodata	&	0.004	\\
Cu	&	0.039	&	0.007	&	3	&	\nodata	&	\nodata	&	\nodata	&	\nodata	&	\nodata	&	0.008	\\
Zn	&	0.009	&	0.012	&	2	&	\nodata	&	\nodata	&	\nodata	&	\nodata	&	\nodata	&	0.012	\\

\enddata

\tablecomments{Abundances of V, Mn, Co, and Cu account for HFS.}

\tablenotetext{a}{NLTE abundances are reported for these elements.  LTE abundances for HIP~102152 are [Cr/H] = -0.019, [Mn/H] = -0.044, and [Co/H] = -0.049.  LTE abundances for 18~Sco are [Cr/H] = 0.060,  [Mn/H] = 0.045, and [Co/H] = 0.028.}

\end{deluxetable}


\begin{thebibliography}{}

\bibitem[Alexander et al.(2001)]{2001Sci...293...64A} Alexander, C.~M.~O., 
Boss, A.~P., \& Carlson, R.~W.\ 2001, Science, 293, 64
\bibitem[Asplund et 
al.(2009)]{2009ARA&A..47..481A} Asplund, M., Grevesse, N., Sauval, A.~J., \& Scott, P.\ 2009, \araa, 47, 481 
\bibitem[Baumann et 
al.(2010)]{2010A&A...519A..87B} Baumann, P., Ram{\'{\i}}rez, I., Mel{\'e}ndez, J., Asplund, M., \& Lind, K.\ 2010, \aap, 519, A87 
\bibitem[Bergemann 
\& Cescutti(2010)]{2010A&A...522A...9B} Bergemann, M., \& Cescutti, G.\ 2010, \aap, 522, A9 
\bibitem[Castelli 
\& Kurucz(2004)]{2004astro.ph..5087C} Castelli, F., \& Kurucz, R.~L.\ 2004, arXiv:astro-ph/0405087 
\bibitem[Chambers(2010)]{2010ApJ...724...92C} Chambers, J.~E.\ 2010, \apj, 
724, 92 
\bibitem[Charbonnel 
\& Talon(2005)]{2005Sci...309.2189C} Charbonnel, C., \& Talon, S.\ 2005, Science, 309, 2189 
\bibitem[Ciesla(2008)]{2008M&PS...43..639C} Ciesla, F.~J.\ 2008, Meteoritics and Planetary Science, 43, 639 
\bibitem[Dekker et al.(2000)]{2000SPIE.4008..534D} Dekker, H., D'Odorico, 
S., Kaufer, A., Delabre, B., \& Kotzlowski, H.\ 2000, \procspie, 4008, 534 
\bibitem[Denissenkov(2010)]{2010ApJ...719...28D} Denissenkov, P.~A.\ 2010, 
\apj, 719, 28 
\bibitem[Do Nascimento et 
al.(2009)]{2009A&A...501..687D} Do Nascimento, J.~D., Jr., Castro, M., Mel{\'e}ndez, J., et al.\ 2009, \aap, 501, 687 
\bibitem[do Nascimento et al.(2013)]{2013ApJ...771L..31D} do Nascimento, 
J.-D., Jr., Takeda, Y., Mel{\'e}ndez, J., et al.\ 2013, \apjl, 771, L31 
\bibitem[Ghezzi et al.(2010)]{2010ApJ...724..154G} Ghezzi, L., Cunha, K., 
Smith, V.~V., \& de la Reza, R.\ 2010, \apj, 724, 154 
\bibitem[Gonzalez et al.(2010a)]{2010MNRAS.407..314G} Gonzalez, G., Carlson, 
M.~K., \& Tobin, R.~W.\ 2010a, \mnras, 407, 314 
\bibitem[Gonzalez et al.(2010b)]{2010MNRAS.403.1368G} Gonzalez, G., Carlson, 
M.~K., \& Tobin, R.~W.\ 2010b, \mnras, 403, 1368 
\bibitem[Israelian et al.(2009)]{2009Natur.462..189I} Israelian, G., 
Delgado Mena, E., Santos, N.~C., et al.\ 2009, \nat, 462, 189 
\bibitem[Kim et al.(2002)]{2002ApJS..143..499K} Kim, Y.-C., Demarque, P., 
Yi, S.~K., \& Alexander, D.~R.\ 2002, \apjs, 143, 499 
\bibitem[Lambert 
\& Reddy(2004)]{2004MNRAS.349..757L} Lambert, D.~L., \& Reddy, B.~E.\ 2004, \mnras, 349, 757 
\bibitem[Lind et 
al.(2009)]{2009A&A...503..541L} Lind, K., Asplund, M., \& Barklem, P.~S.\ 2009, \aap, 503, 541 
\bibitem[Mel{\'e}ndez et al.(2009)]{2009ApJ...704L..66M} Mel{\'e}ndez, J., 
Asplund, M., Gustafsson, B., \& Yong, D.\ 2009, \apjl, 704, L66 
\bibitem[Mel{\'e}ndez et 
al.(2010)]{2010Ap&SS.328..193M} Mel{\'e}ndez, J., Ram{\'{\i}}rez, I., Casagrande, L., et al.\ 2010, \apss, 328, 193 
\bibitem[Mel{\'e}ndez et 
al.(2012)]{2012A&A...543A..29M} Mel{\'e}ndez, J., Bergemann, M., Cohen, J.~G., et al.\ 2012, \aap, 543, A29 
\bibitem[Nordlund(2009)]{2009arXiv0908.3479N} Nordlund, A.\ 2009, 
arXiv:0908.3479 
\bibitem[{\"O}nehag et 
al.(2011)]{2011A&A...528A..85O} {\"O}nehag, A., Korn, A., Gustafsson, B., Stempels, E., \& Vandenberg, D.~A.\ 2011, \aap, 528, A85 
\bibitem[Ram{\'{\i}}rez et 
al.(2009)]{2009A&A...508L..17R} Ram{\'{\i}}rez, I., Mel{\'e}ndez, J., \& Asplund, M.\ 2009, \aap, 508, L17 
\bibitem[Ram{\'{\i}}rez et 
al.(2010)]{2010A&A...521A..33R} Ram{\'{\i}}rez, I., Asplund, M., Baumann, P., Mel{\'e}ndez, J., \& Bensby, T.\ 2010, \aap, 521, A33 
\bibitem[Ram{\'{\i}}rez et al.(2011)]{2011ApJ...740...76R} Ram{\'{\i}}rez, 
I., Mel{\'e}ndez, J., Cornejo, D., Roederer, I.~U., 
\& Fish, J.~R.\ 2011, \apj, 740, 76 
\bibitem[Ram{\'{\i}}rez et al.(2012)]{2012ApJ...756...46R} Ram{\'{\i}}rez, 
I., Fish, J.~R., Lambert, D.~L., \& Allende Prieto, C.\ 2012, \apj, 756, 46 
\bibitem[Schuler et al.(2011)]{2011ApJ...737L..32S} Schuler, S.~C., Cunha, 
K., Smith, V.~V., et al.\ 2011, \apjl, 737, L32 
\bibitem[Sneden(1973)]{1973PhDT.......180S} Sneden, C.~A.\ 1973, 
Ph.D.~Thesis,  
\bibitem[Sousa et 
al.(2007)]{2007A&A...469..783S} Sousa, S.~G., Santos, N.~C., Israelian, G., Mayor, M., \& Monteiro, M.~J.~P.~F.~G.\ 2007, \aap, 469, 783 
\bibitem[Takeda et 
al.(2007)]{2007A&A...468..663T} Takeda, Y., Kawanomoto, S., Honda, S., Ando, H., \& Sakurai, T.\ 2007, \aap, 468, 663
\bibitem[Takeda 
\& Tajitsu(2009)]{2009PASJ...61..471T} Takeda, Y., \& Tajitsu, A.\ 2009, \pasj, 61, 471 
\bibitem[Xiong 
\& Deng(2009)]{2009MNRAS.395.2013X} Xiong, D.~R., \& Deng, L.\ 2009, \mnras, 395, 2013 



\end{thebibliography}
\end{document}